\documentclass[prl,showpacs,twocolumn]{revtex4}

\usepackage{graphicx}

\usepackage{amsmath}
\usepackage{amsfonts}
\usepackage{amssymb}
\usepackage{bm}

\begin{document}

\title{$2\delta$-Kicked Quantum Rotors: Localization and `Critical' Statistics}

\author{C.E.~Creffield, G.~Hur and T.S.~Monteiro}

\affiliation{Department of Physics and Astronomy, University College London,
Gower Street, London WC1E 6BT, United Kingdom}

\date{\today}

\begin{abstract}
The quantum dynamics of atoms subjected to pairs of 
closely-spaced $\delta$-kicks from optical potentials are shown to be
quite different from the well-known
paradigm of quantum chaos, the singly-$\delta$-kicked system.
We find the unitary matrix has a new oscillating band structure
corresponding to a cellular structure of phase-space and observe a spectral signature of
a localization-delocalization transition from one cell to several.
We find that the eigenstates
have localization lengths which scale with a fractional
power $L \sim \hbar^{-.75}$ and obtain  a regime of near-linear spectral variances 
which approximate the
`critical statistics' relation 
$\Sigma_2(L) \simeq \chi L \approx \frac{1}{2}(1-\nu) L$, 
where $\nu \approx 0.75$ is related to the fractal classical phase-space 
structure. The origin of the  $\nu \approx 0.75$ exponent is analyzed. 
\end{abstract}


\maketitle

The $\delta$-kicked quantum rotor (QKR)
is one of the most studied paradigms of quantum chaos.
Its implementation using cold atoms in
optical lattices \cite{Raizen} provided a convincing demonstration of
a range of  quantum chaos phenomena including
Dynamical Localization \cite{Fish}, the quantum suppression
of chaotic diffusion. 
Recently, an experimental study \cite{Jones} of 
cesium atoms subjected to {\em pairs} of $\delta$-kicks (2$\delta$-KR)
showed surprisingly different behavior.
The classical phase-space is chaotic but
is made up fast diffusing regions which are partly separated by
slow-diffusing `trapping regions', where the classical trajectories stick;
the classical analysis revealed a regime of anomalous 
diffusion corresponding to long-lived correlations between kicks.

Further details are given in \cite{Web} but we show here that the
2$\delta$-KR has some unexpected {\em quantum} properties.
We show that there is a cellular phase-space structure which
arises from a novel oscillatory band structure of the corresponding
unitary matrix. One consequence is a new type of localization-delocalization
transition not seen in the QKR, where states delocalize from single
to multiple-cell occupancy; we show it has a clear spectral signature.
We have also found scaling behavior of the localization lengths
associated with a {\em fractional} exponent, i.e. $L \sim \hbar^{-0.75}$,
whereas for the well-studied QKR, $L \sim \hbar^{-1}$.
A similar exponent is found for the decay of return probabilities in the
trapping regions, $P(t) \sim t^{-0.75}$.
We argue that the exponent $0.75$ corresponds closely to
the value obtained for the dominant exponent of the 
golden ratio cantorus \cite{Fish2,Maitra}. 
We show that the spectral fluctuations (both the nearest-neighbor
statistics (NNS) and spectral
variances) show important differences with the QKR in regimes 
where the delocalization of eigenstates is hindered by cantori
bordering the cells. We find a regime 
approximating the form found in `critical statistics': the number variances
of the spectra are linear $\Sigma_2(L) \simeq \chi L$ for $L \gg 1$,  
where $\chi \simeq 1/2(1-\nu) < 1$ and  $\nu \simeq 0.75$.
 
The term `critical statistics' arose originally in relation to the 
Metal Insulator Transition (MIT) in systems with 
disorder \cite{Shapiro,Chalker}.
A new universal form of the distribution of nearest-neighbor
eigenvalue spacings, termed `semi-Poisson', $P(s)\sim s \exp{-2s}$
was associated with the MIT \cite{Shapiro}.
For critical statistics a very interesting connection has been 
established between 
the multifractal characteristics of the wavefunctions and those of the
spectral fluctuations \cite{Chalker}: the number variances
of the spectra are linear $\Sigma_2(L) \simeq \chi L$ for $L \gg 1$. 
The slope, $\chi \simeq 1/2(1-D_2/D) < 1$, was shown to be related to a fractal
dimension $D_2$ obtained from the second moment of the wavefunction and 
to $D$, the spatial dimension of the system. For integrable
dynamics, in contrast, $\Sigma_2(L) = L$ while 
for a GOE, $\Sigma_2(L) \sim Ln(L)$.
There is much current interest in so-called `critical' statistics
in non-KAM billiards (typically
systems where the dynamics would be integrable were it not for a discontinuity
in the potential) \cite{Garcia}, which show multifractal scalings and linear
variances related to $D_2$. Below we apply the term `critical statistics'
in this broader sense, rather than the MIT critical point.
Multifractal behavior has been demonstrated for Cantor spectra
\cite{Ketz} where the level density itself is not smooth.
However, until now, critical statistics have not been seen -- and were not
thought to be relevant to -- KAM systems. These are systems, ubiquitous in
many areas of physics, where the transition to chaos
as a perturbing parameter is increased is quite gradual.

\begin{figure}[b]
\includegraphics*[width=0.125\textwidth,angle=-90,clip=true]{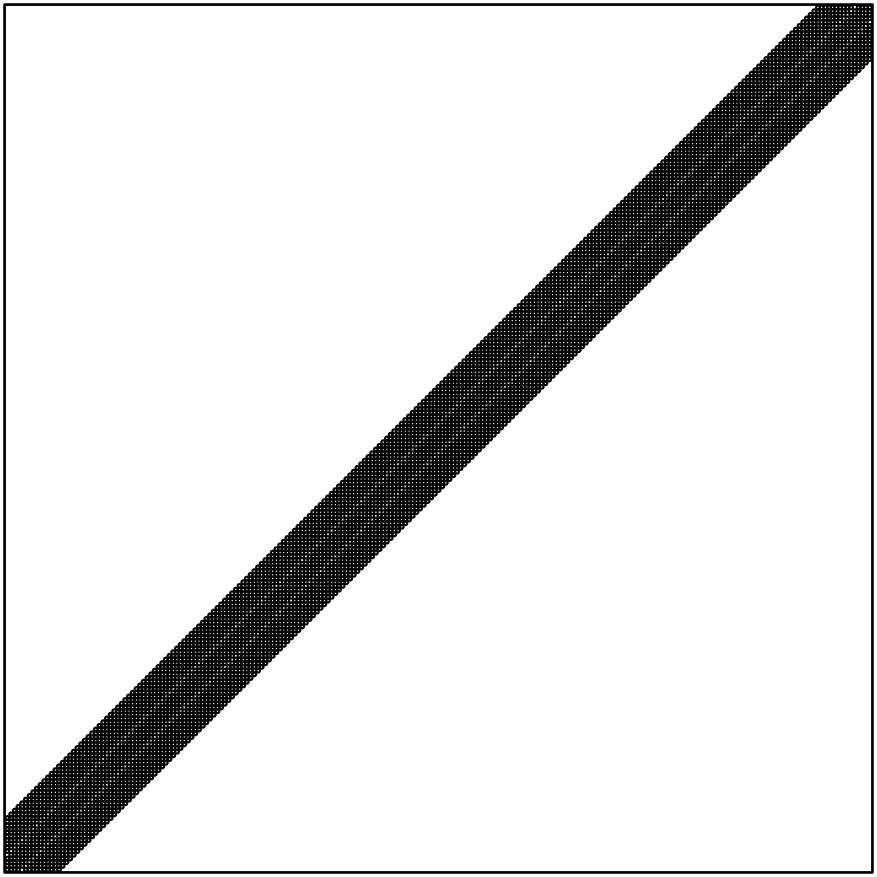}
\hspace*{5 mm}
\includegraphics[width=0.125\textwidth,angle=-90,clip=true]{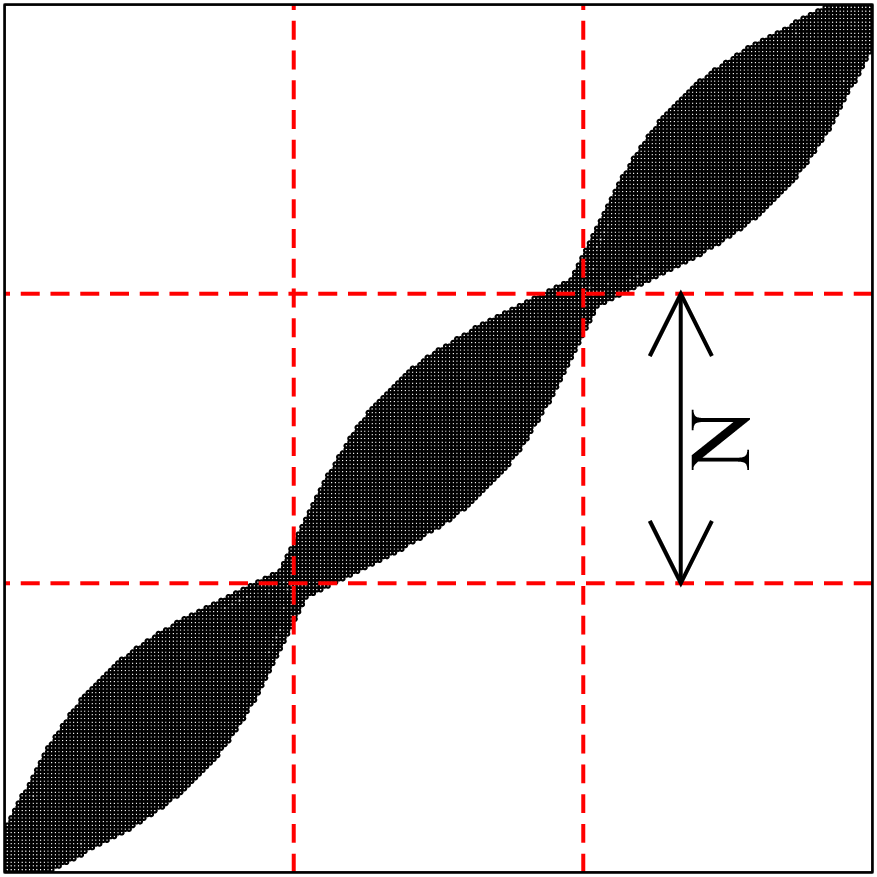}
\caption{Left: Structure of time-evolution matrix $U(T,0)$, 
for the Quantum Kicked Rotor (QKR),
in a basis of momentum states,
exemplifying the typical band-structure of a Band Random Matrix (BRM). 
Right: $U(T,0)$ for our system, the 2$\delta$-KR, showing the new form with
oscillating bandwidth. Before delocalization, 
eigenstates are confined within a single `momentum cell' of dimension $N$.}
\label{Fig1}
\end{figure}
                                                                                                                                                             
The  Hamiltonian of the 2$\delta$-KR is
$H(x,p)= \frac{p^2}{2} + K \cos x \sum_n \delta(t-nT) + \delta(t-nT+\epsilon)$;
there is a short time interval $\epsilon$ between the kicks in each pair
and a much longer time $\simeq T$ between the pairs themselves.
In experiments $\epsilon \sim 0.01 - 0.1 \ll T$ and $ \hbar  \simeq 2 \to 1/4$
in the usual re-scaled units \cite{Jones}.

A study of the spectral fluctuations of a time-periodic system involves a
study of the eigenstates and eigenvalues
of the one-period time-evolution operator $U(T,0)$. 
For the QKR, the matrix representation, in an angular 
momentum basis $|\ l \rangle$, has elements
$U_{lm}= U_l^{free}. \ U_{lm}^{kick}=\exp {-i l^2T\hbar/2}  \ . \ 
 J_{l-m}(\frac{K}{\hbar})$.
The `kick' terms, $J_{l-m}(\frac{K}{\hbar})$
are Bessel functions and
give the matrix the banded form illustrated in Fig.\ref{Fig1}(a).
Since $J_{l-m}(x) \simeq 0$ if $|l-m| >x$ we define the 
usual bandwidth $b = K/\hbar$. 
The resulting statistics are approximated by those of Band Random Matrix Theory
(BRMT) \cite{Izraelev} rather than of RMT: i.e.
if the dimension of the $U(T,0)$
matrix is $N_{tot}$, the statistics are {\em Poissonian} for $N_{tot}\gg b$;
the eigenstates of the BRM are exponentially localized in $l$, 
with a localization length in momentum ($p=l\hbar$) which equals 
$L_p \sim K^2/\hbar$, so states separated
in $p$ by $\gg L_p$ will be largely uncorrelated.

For the 2$\delta$-KR, the corresponding matrix elements are:
\begin{equation}
U_{lm}=\\
 e^{-i l^2 (T-\epsilon) \hbar/2} \ . \ 
\sum_k J_{l-k}\left(\frac{K}{\hbar}\right)
J_{k-m}\left(\frac{K}{\hbar}\right)
 e^{-i \frac{k^2}{2} \hbar\epsilon} .
\end{equation}
As the $U(T,0)$ matrix is quite insensitive to $(T-\epsilon)$ 
the quantum dynamics largely depends only on
two scaled parameters $K_\epsilon= K\epsilon$ 
and $\hbar_{\epsilon}= \hbar\epsilon$, rather than
on $K,\epsilon$ and $\hbar$ independently: the remainder of the matrix is
invariant if  $K_\epsilon$ and $\hbar_{\epsilon}$
are kept constant.

An analytical form for the bandwidth of $U(T,0)$ was obtained 
in \cite{Web}: the bandwidth oscillates sinusoidally between
a maximum value $b_{max}= 2K/\hbar$ for angular momenta 
$l \simeq 2n\pi/(\hbar\epsilon)$
and a minimum value $b_{min} \approx 0$ for 
$l \simeq (2n+1)\pi/(\hbar\epsilon)$.
These minima correspond to the trapping momenta 
$p= l\hbar \simeq (2n+1)\pi/\epsilon$ 
seen in experiment. The corresponding band-structure of $U$ is
illustrated in Fig.\ref{Fig1}(b): the
band oscillates, and $U$ is approximately partitioned into sub-matrices
of dimension $N=\frac{2\pi}{\epsilon \hbar}$ corresponding to 
separate momentum cells.

The key to our work is our ability to vary the {\em transport} between the cells
(by opening/closing the classical fractal `gates' between them) separately
from the degree of {\em filling} of each individual cell. 
We begin by introducing a `filling factor' $R$ where
\begin{equation}
R=\frac{K^2}{N \hbar^2} = \frac{K^2_\epsilon}{2 \pi \hbar_{\epsilon}}
\label{eq2}
\end{equation}
measures the degree of filling of a cell by a typical state in the 
absence of confinement.
Clearly, if there is no transport between cells the states simply 
fill the cell uniformly
and even if $R \gg 1$, the localization lengths $L_p \sim  N\hbar$.
We begin by defining a {\em localized} limit, where $R \ll 1$ and 
$L_p \ll  N\hbar \sim \frac{2\pi}{\epsilon} $; typical states are 
insensitive to the boundary conditions of the cell and this limit 
is Poissonian. At the other extreme, if
we allow strong coupling between cells, we move to an opposite limit as an
increasing proportion of eigenstates become delocalized over several cells.

We now investigate the transport. A classical analysis \cite{Web,Mischa} shows 
that if we take $K \epsilon \ll 1$ and expand initial momenta 
of the $j-th$ trajectory about the trapping values
$p_j = (2n+1)\pi/\epsilon + \delta p_j$, we can show that much of the trapping
region is given by a classical map 
quite similar to the well-known Standard Map:
\begin{eqnarray}
p_{j+2} &\simeq& p_{j} - K^2 \frac{\epsilon}{2} \ \sin 2 x_{j} - K \delta p^j \  \epsilon \cos x_{j}\label{eq13} \\
x_{j+2} &\simeq& x_{j} + p_{j+2} T . 
\label{eq13a}
\end{eqnarray}

Over much of the trapping region, the second term in Eq.\ref{eq13}
is dominant, for parameter regimes of interest.
Then the kick impulse has a $\pi/2$ phase relative to the full map 
(the 2-kick map gives a pair of $V'(x) = -\sin x$ type impulses) 
and a momentum dependent effective kick strength 
$K'= K \epsilon \delta p_j $. A detailed study of classical
phase-space  \cite{Mischa} shows that at low $K'$ the resonance structure is locally
quite similar to 
the Standard Map. Hence, in the regime $ K \epsilon \delta p_j \sim 1$ 
we expect the
`golden ratio' cantori, which result from the last invariant manifold
of a KAM system, to provide the strongest barrier to transport \cite{Ott}. 
Though we want $K \epsilon$ to be small, if $K \epsilon < 0.1$
phase space becomes too regular. Hence here we find that the regime of interest is within the interval $0.1 \leq K_\epsilon \leq 0.7$.

\begin{figure}[htb]
\includegraphics[height=2.25in]{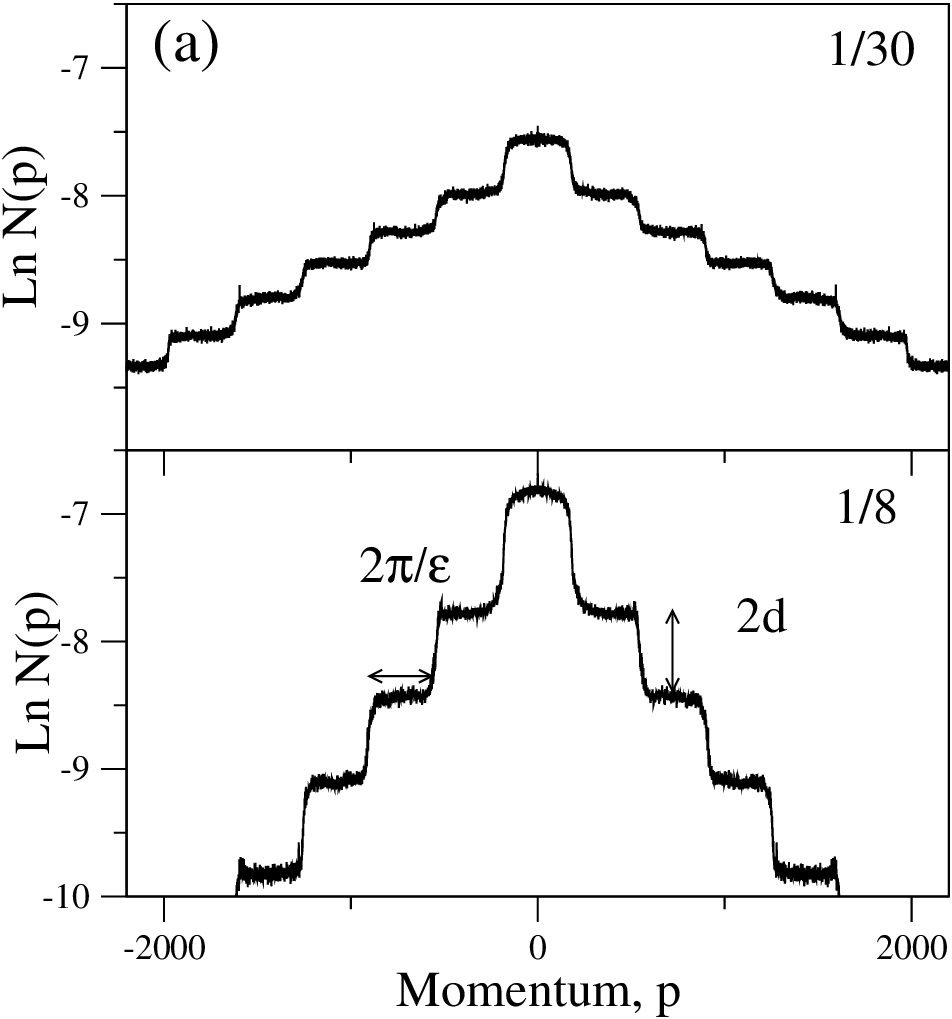}
\includegraphics[height=2.in]{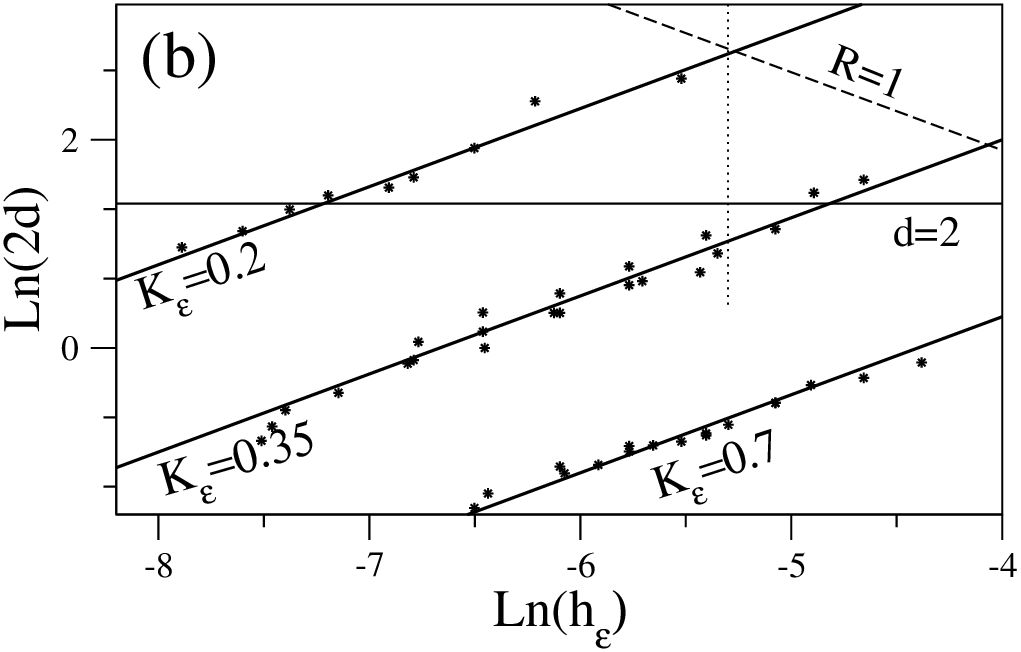}
\caption { {\bf(a)} Final ($t \to \infty$) momentum distributions, $N(p)$, 
(slightly smoothed) for quantum wavepackets of the 2$\delta$-QKR
for $K=20$, $\epsilon=0.0175$ and $\hbar=1/8$ and $ 1/30$ respectively.
$N(p)$ for both the eigenstates and wavepackets shows a long-range
`staircase' form which on average follows the exponential
$N(p) \sim \exp{-2(p-{\overline{p}})/L_{exp}}$ where 
$L_{exp}/2 = 2\pi/\hbar\epsilon$;  
the $\hbar$-dependence of $L_{exp}$ is  determined
by the drop in probability, $d$, at each step.
{\bf(b)} shows that $Ln(2d)$ plotted against $Ln (\hbar \epsilon)$ 
lies on straight lines of invariant $K_{\epsilon} = K \epsilon$, 
with constant slope $0.75$. Hence $d \propto (\hbar \epsilon)^{0.75}$
and $L_{exp} \propto \hbar^{-0.75}$ -- in contrast to the well-known
QKR result $L_{exp} \propto \hbar^{-1}$.
The $R=1$ border is shown: for $R>1$ (below the line) eigenstates
fill much of a single cell. The {\em delocalization} border is shown 
at $d \simeq 2$: for $d>2$ (above the line), over $98\%$ of the probability 
of typical eigenstates is 
confined to a single cell; for $d<2$, eigenstates begin to occupy 
multiple cells.  Statistics  are presented
later in Fig.3 for points corresponding to the dotted line. }
\label{Fig2}
\end{figure}

We investigated the corresponding quantum transport 
by evolving a set of wavepackets $\Phi(p,t)$ in time, 
(where $\Phi(p,t=0) = \delta(p)$)  for a range
of $K_\epsilon $ and $\hbar_\epsilon$, until the momentum
spreading is arrested by dynamical localization at $t \simeq t_H$. 
The resulting probability
distributions $|\Phi(p,t \gg t_H)|^2= N(p)$ have a characteristic `staircase' structure, shown in Fig.\ref{Fig2}(a).
At each step there is a steep drop in probability:
\begin{equation} 
N(p)_+= e^{-{2d}}N(p)_-  
\end{equation}
(where $N(p)_\pm$ represent probabilities before(-) and
after(+) the step)
concentrated over the trapping region 
($\sim 1/6$ of a cell in every case \cite{Web}).
The staircase tracks an exponential envelope
$N(p) \sim \exp-{2|p|/L_{exp}}$, where 
$L_{exp}=\frac{\pi}{\epsilon d}$. We average over
several steps, to obtain $d$ as a function of $K_\epsilon$ and
$\hbar_\epsilon$. In Fig.\ref{Fig2}(b) we show that, quite accurately,
$d \propto \hbar_\epsilon^{0.75}/f(K_\epsilon)$ where $f(K_\epsilon)$ is
some function of the scaled kick-strength, $K_\epsilon$. We estimate
\begin{equation}
d \approx \frac{3.5\hbar_\epsilon^{0.75}} {K_\epsilon^3} .
\label{eq4}
\end{equation}
This defines our transport parameter, $d$, and complements the `filling factor'
$R$. The inner steps of the staircase have been seen in the
momentum distributions of atoms in optical lattices 
\cite{Christensen,Jones}; so,
though existing data is
not in the critical regime, in principle the form of $d$ is 
experimentally verifiable.

We are unaware of another KAM system where a single power-law exponent 
is so dominant. Typically, power-law behavior is associated 
with mixed phase space behavior, where many
competing exponents are found \cite{Ketz}. We note also that the 
value $0.75$ coincides closely with one of the scaling exponents found 
in \cite{Fish2} for the golden ratio cantori: these were
$\sigma \approx 0.65$ (in the most unstable part of the cantori) and $\sigma \approx 0.76$ (most stable regions). 
There have been previous studies of
transport in a region near golden ratio cantori. These have found 
$L \sim \hbar^{+0.66}$ \cite{Geisel} 
but only in a momentum band region `local' to the cantori.
We note that the $L \sim \hbar^{+\sigma}$ dependence is associated with 
the physical process termed `retunnelling' \cite{Maitra} associated 
with cantori which are classically `open'
but for which $\hbar$ is too large to permit free quantum transport. 
An abrupt
change to an $ L \sim \hbar^{-\sigma}$ is observed when the cantori `open'
for quantum transport and $L$ is determined instead by localization. 
It has been argued that the reason all previous studies have 
found $L \sim \hbar^{+0.66}$ 
\cite{Maitra} is that the retunnelling transport favors 
the unstable direction. To our knowledge ours is the only example 
corresponding to the dominant exponent of
the golden ratio cantori; we attribute this to the fact 
that we are always in a dynamical
localization regime, and localization will select the most stable parts of the
fractal cantori regions, where at low $K$, elliptic fixed points are found.

We can now investigate the statistics as a function of the filling 
factor $R$ and the inter-cell transport parameter $d$. 
Full details are given in \cite{Web}, but
in brief: we considered two types of boundary conditions (BCs).
(1) Periodic BCs, i.e. solving the problem on a `torus' in momentum space,
a well-known procedure for the QKR \cite{Izraelev}. 
(2) Open BCs, where we diagonalize 
$U(T,0)$ with $N_{tot}=10,000$,
but $N \approx 1000$; we then assigned the $i-th$ eigenstate 
to the $n-th$ cell if 
$(2n+1)\pi/\epsilon \leq \langle p_i \rangle \leq (2n+3)\pi/\epsilon$ 
and calculated statistics for
each cell. In both cases we averaged over $\approx 20$ cells to improve 
significance. For periodic BCs, eigenstates cannot escape from a single cell 
of width $N\hbar$.
For open BCs, however, they can delocalize onto neighboring cells.

\begin{figure}[htb]
\includegraphics[height=1.5in,width=3.in]{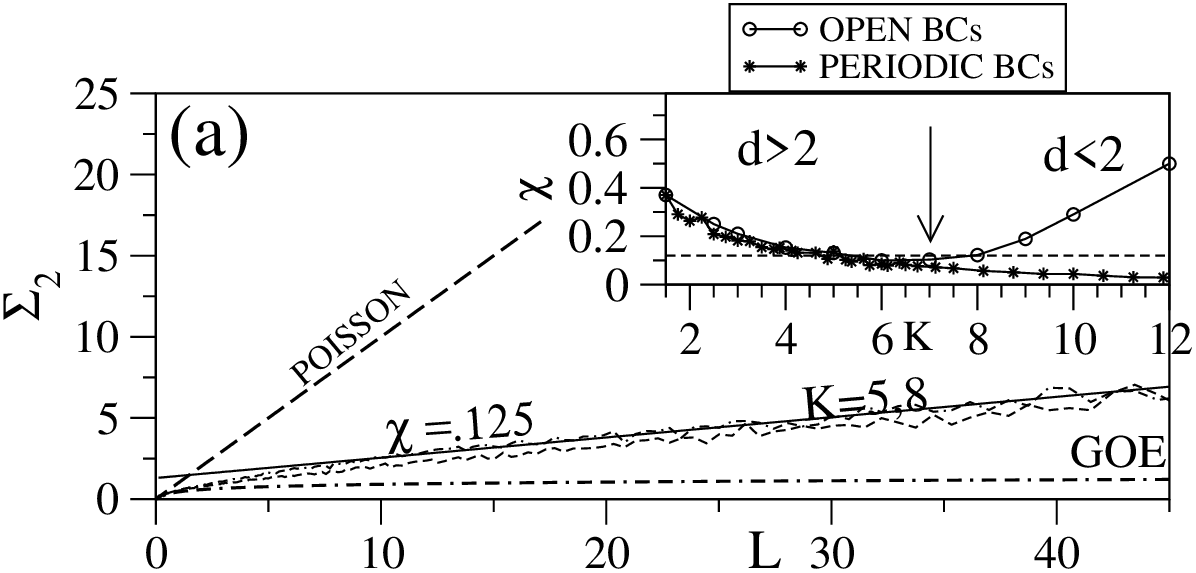}
\vspace*{10 mm}
\includegraphics[height=1.5in,width=3.in]{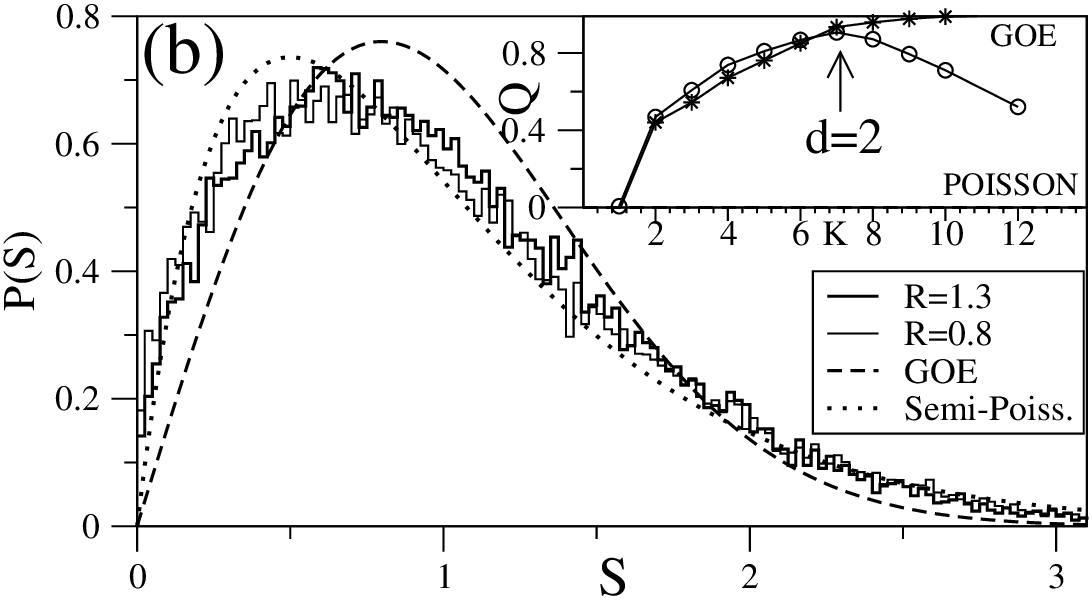}
\caption {{\bf(a)} shows the  $\Sigma_2(L)$ statistics for the
$2\delta-KR$. Approximately linear variances, with slope $\chi \approx 0.125$
are found in the intermediate regime. Solid line indicates the line
$\Sigma_2(L) = 0.125 L + c$  corresponding to the `critical statistics' slope
$\chi \approx 1/2(1-\nu)$ where $\nu \simeq 0.75$ is the fractional 
exponent obtained
previously. {\bf(b)} shows corresponding nearest neighbor statistics
for $R=0.8$ and $R=1.3$ of an intermediate form, which is compared with the
Semi-Poisson and GOE forms.
The insets show the effect of the boundary conditions. Before
delocalization ($d>2$) the results are insensitive to boundary conditions.
At delocalization, if the eigenstates are confined to a single cell 
(periodic BCs, points indicated by asterisks) the
statistics make a transition to GOE; if they can delocalize (open BCs, points
indicated by circles), 
the statistics tend to Poissonian behavior.
The arrows indicate the $d=2$ border.}
\label{Fig3}
\end{figure}

In Fig.\ref{Fig3}(a), the $\Sigma_2(L)$ statistics are presented. 
These represent the
variances in the spectral number density,
$\Sigma_2(L)= \langle L^2  \rangle - \langle L \rangle^2$, where we consider a
stretch of the spectrum with an average $\langle L \rangle$ levels. 
A fit to the best straight line
in the range $L \simeq 5-40$ yields an estimate of the slope $\chi$.
In Fig.\ref{Fig3}(b) we show the
nearest neighbor $P(S)$ statistics. We quantify the deviation of $P(S)$ from 
$P_{P}(S)$ and $P_{GOE}(S)$,
its Poisson and GOE limits respectively,
with a parameter $Q$ \cite{Casati2}:
\begin{equation}
1-Q= \frac{ \int_0^{S_0} ( P(S') - P_{GOE}(S') dS')}
{\int_0^{S_0} (P_{P}(S') - P_{GOE}(S'))dS'} .
\end{equation}
Hence $Q=0$ indicates a Poisson distribution, while $Q=1$
signals a GOE distribution. We take $S_0=0.3$.

The results of Fig.\ref{Fig3} demonstrate that the statistics are 
not too sensitive
to boundary conditions for $d>2$ (see also \cite{Web}). However,
for $d \leq 2$, while the states with periodic BCs 
(effectively restricted to a matrix of dimension $N$) 
move gradually to the GOE limit,  for the open BCs, at delocalization
 the statistics tend back to the Poisson limit. This initially surprising behavior occurs
because, after delocalization, one finds increasing numbers of states for
which $\langle p_i \rangle$ assigns them to the $n-th$ cell, but for
which much of the state's probability is actually found in neighboring cells
\cite{Web}. Hence
states within each spectrum of $N$ eigenvalues
become progressively uncorrelated. This apparent failure of the
procedure of assigning states to a given cell, in fact provides a
rather good `marker'
for the onset of delocalization, and
yields a clear `turning point' in both the NNS 
and $\Sigma_2(L)$ behavior. A detailed study of the multi-cell regime
remains to be undertaken; models of the statistics for chaotic systems
with non-uniform rate of exploration of phase-space \cite{Bohigas} may
be relevant here.

Of further interest here is an intermediate regime, found for both boundary
conditions for $R \approx 1$ and $d > 2$. Over a wide range of parameters
and different cell sizes \cite{Web}, we find approximately 
linear variances, for $L \gg 1$ and $L \ll N$, with slope $\chi \approx 0.125$.
The inset of Fig.\ref{Fig3}(a) plots the values of $\chi$ calculated along 
the vertical dotted line
of Fig\ref{Fig2}(b) for periodic and open boundary conditions. We note that
a study of the decay of return probabilities 
$P(t)= |\langle \psi(t=0)| \psi(t)| \rangle^2 \sim t^{-0.75}$
for wavepackets started in the trapping regions (see \cite{Web}). 
The value of $\chi \approx 0.125$ corresponds  to the value 
which would be obtained from the MIT relation, 
$\chi \approx 1/2(1-D_2)$ with, if $D_2 \approx 0.75$; in the MIT,
return probabilities with $P(t) \sim t^{-D_2}$ were similarly found.
We suggest that this represents a KAM analogue of behavior associated with
`critical statistics'.

In summary, we have shown that the behavior of the 2$\delta$-kicked system
is rather different from the standard QKR. We have identified
a  spectral signature of the novel localization-delocalization
transition that the system exhibits. We have also identified signatures
of the fractal phase-space structure of the cell borders.
 The trapping regions may have applications
in atom optics experiments, as a means of manipulating the momentum 
distribution of the atomic cloud.

This work was supported by the EPSRC. We thank Shmuel Fishman 
and Antonio Garcia-Garcia for helpful comments and advice.

\end{document}